\documentclass[a4paper,10pt]{IEEEtran}

\usepackage{amsmath}
\usepackage{amssymb}
\usepackage{amsfonts}
\usepackage{graphicx}

\usepackage{xcolor}

\usepackage{physics}

\usepackage{setspace}

\usepackage{cite}

\title{Learning Quantum Entanglement Distillation with Noisy Classical Communications}

\author{
\IEEEauthorblockN{Hari Hara Suthan Chittoor,~\IEEEmembership{Member,~IEEE } and Osvaldo Simeone,~\IEEEmembership{Fellow,~IEEE }}\vspace{-0.5cm}}

\begin{document}

\maketitle

\thispagestyle{empty}	

\pagestyle{empty}


\let\thefootnote\relax\footnotetext
{The authors are with King’s Communications, Learning, and Information
Processing (KCLIP) lab at the Department of Engineering of Kings College
London, UK (emails: hari.hara@kcl.ac.uk, osvaldo.simeone@kcl.ac.uk).
The authors have received funding from  the European Research Council (ERC) under the European Union's Horizon 2020 Research and Innovation Programme (Grant Agreement No. 725731).

}

\begin{abstract}

Quantum networking relies on the management and exploitation of entanglement. Practical sources of entangled qubits are imperfect, producing mixed quantum state with reduced fidelity with respect to ideal Bell pairs. Therefore, an important primitive for quantum networking is entanglement distillation, whose goal is to enhance the fidelity of entangled qubits through local operations and classical communication (LOCC). Existing distillation protocols assume the availability of ideal, noiseless, communication channels. In this paper, we study the case in which communication takes place over noisy binary symmetric channels. We propose to implement local processing through parameterized quantum circuits (PQCs) that are optimized to maximize the average fidelity, while accounting for communication errors.  The introduced approach, Noise Aware-LOCCNet (NA-LOCCNet), is shown to have significant advantages over existing protocols designed for noiseless communications.

\end{abstract}



\vspace{-0.2cm}

\section{Introduction}

\begin{figure}[htbp]
\centering
\includegraphics[height=2.9in]{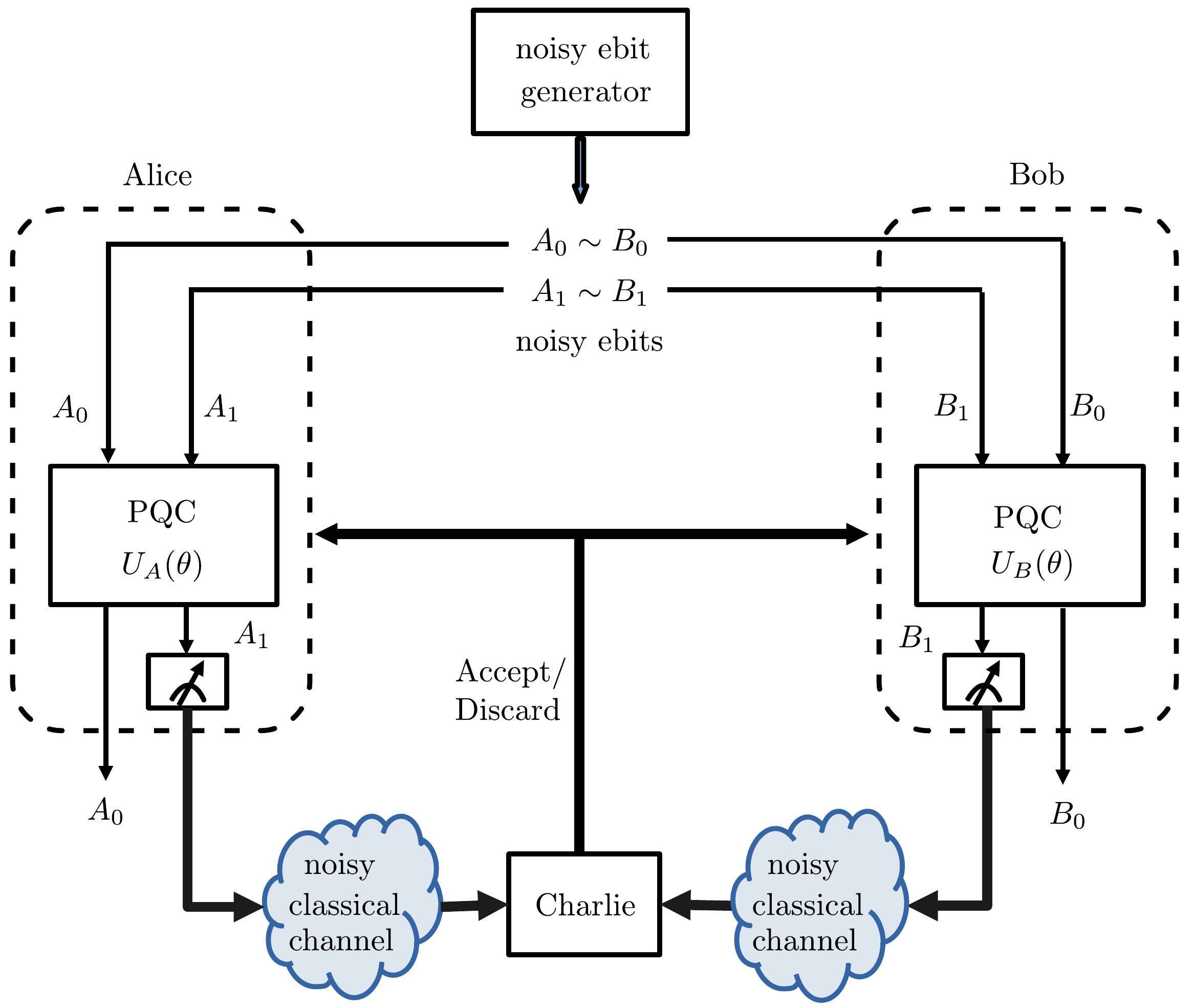}
\caption{Entanglement distillation at two quantum-enabled devices (Alice and Bob) aided by a noisy classical communication channel to a third party (Charlie).}
\label{fig: Alice Bob Charlie}
\end{figure}

Quantum networking, and with it the quantum Internet, rely on the management and exploitation of entanglement \cite{Book_Quantum_internet_second_quantum_revolution_2021,Book_quantum_networking_Meter,Quantum_Internet_IEEE_Network_2020}. In fact, entangled qubits enable fundamental quantum communication primitives such as teleportation and superdense coding \cite{Book_quantum_computing_quantum_information_Nielsen_chuang_2010,Book_Quantum_Information_Theory_wilde_2013}. Practical sources of entangled qubits, such as single-photon detection \cite{single_photon_sources_and_detectors,Sstate_ref_paper2}, are imperfect, producing mixed states with reduced fidelity as compared to ideal, fully entangled, Bell pairs. In order to enhance the fidelity of entangled qubits available at distributed parties, \textit{entanglement distillation} protocols leverage \textit{local operations and classical communication (LOCC)}. While existing solutions assume ideal classical communications, this paper studies the case in which communications between the parties holding imperfectly entangled qubits are noisy. As illustrated in Fig. \ref{fig: Alice Bob Charlie}, to address this more challenging scenario, we propose the use of quantum machine learning via \textit{parameterized quantum circuits (PQCs)} \cite{book_machine_learning_with_quantum_computers_Maria_Francesco,Book_Osvaldo_quantum_machine_learning_for_engineers}.

In entanglement distillation protocols, a source produces a number of imperfectly entangled qubit pairs. Each qubit of a pair is made available at one of two parties, conventionally referred to as Alice and Bob. The goal is to leverage LOCC to produce qubit pairs that have a higher degree of fidelity with respect to a fully entangled Bell pair. In the most typical case, Alice and Bob start with two qubit pairs, and output either one qubit pair or a declaration of failure at the end of the process (see Fig. \ref{fig: LOCCNet} and Fig. \ref{fig: proposed NA-LOCCNet}).

        
Traditionally, entanglement distillation protocols have been designed by hand, targeting specific mixed states as the input of the protocol \cite{DISTILLATION_BBPSW,Book_quantum_networking_Meter,DISTILLATION_DEJMPS}. Specific examples include the DEJMPS protocol, which targets the so-called S-state \cite{DISTILLATION_DEJMPS}. These methods rely on local operations via specific unitaries; on the measurement of one qubit at Alice and Bob; and on classical communication of the measurement outputs on a noiseless channel. Based on the measurement outputs, Alice and Bob decide whether to keep the unmeasured pair of qubits or to declare a distillation failure.

Recently, a quantum machine learning framework was introduced in \cite{DISTILLATION_LOCCNet} for the design of LOCC protocols. The approach, termed \textit{LOCCNet}, prescribes the use of PQCs for the local unitaries applied by Alice and Bob. PQCs have been widely investigated in recent years as means to program small-scale noisy quantum computers via classical optimization, with applications ranging from combinatorial optimization to generative modelling \cite{book_machine_learning_with_quantum_computers_Maria_Francesco}. A PQC typically consists of a sequence of one- and two-qubit rotations, whose parameters can be optimized, as well as of fixed entangling gates.

The design of LOCCNet in \cite{DISTILLATION_LOCCNet} assumes ideal, noiseless, classical communications. In contrast, in this paper, we study the case in which communication  takes place over \textit{noisy binary symmetric channels} as seen in Fig. \ref{fig: Alice Bob Charlie}. We propose to optimize a specific PQC architecture (see Fig. \ref{fig: proposed NA-LOCCNet}) with the goal of maximizing the average fidelity when accounting for the randomness caused by communication errors.  The introduced approach, \textit{Noise Aware-LOCCNet (NA-LOCCNet)}, is shown to have significant advantages over existing protocols designed for noiseless communications.

The rest of the paper is organized as follows. In Section \ref{section problem formulation}, we formulate the problem setting and define the performance metrics of interest. In Section \ref{section literature review}, we review the DEJMPS \cite{DISTILLATION_DEJMPS} and LOCCNet \cite{DISTILLATION_LOCCNet} protocols. In Section \ref{section Noisy-LOCCNet}, we present the proposed NA-LOCCNet protocol that addresses settings with noisy classical communications. Experimental results are presented in Section \ref{section experiments}, and Section \ref{section conclusion} concludes the paper.

\textbf{Notations:} The Kronecker product is denoted by $\otimes$; $I_d$ represents the $d \times d$ identity matrix; $M^{\dagger}$ represents the complex conjugate transpose of the matrix $M$; $\mathrm{tr}(M)$ represents trace of the matrix $M$. We adopt standard notations for quantum states, computational basis, and quantum operations \cite{Book_quantum_computing_quantum_information_Nielsen_chuang_2010}.


\section{Problem formulation}
\label{section problem formulation}

In this section, we formulate the problem of entanglement distillation in the presence of a noisy classical communication channel, and we describe the performance metrics of interest.

\subsection{Setting}
\label{subsection setting}

As illustrated in Fig. \ref{fig: Alice Bob Charlie}, we consider a system consisting of two main parties -- Alice and Bob -- aided by a third party -- Charlie. Alice and Bob have local quantum processing capability, while Charlie is not equipped with quantum computing devices. Alice and Bob can communicate to Charlie over a \textit{noisy classical channel}. An imperfect quantum entanglement mechanism generates pairs of noisy entangled qubits, also referred to as \textit{noisy ebits}. One of the qubits of each entangled pair is made available to Alice and the other to Bob. The goal of the system is to improve the average fidelity, defined in Section \ref{subsection performance metrics} and Section \ref{subsection design objective}, of the noisy ebits shared by Alice and Bob through local operations (LO) at Alice and Bob, as well as through classical communication (CC) to Charlie.

The quantum entanglement generator produces $k$ pairs of noisy ebits. The state of each qubit pair is described by a $4 \times 4$ density matrix $\rho_{AB}$. Throughout the paper, we use subscript $A$ to denote the qubits available at Alice, while the subscript $B$ is used for the qubits at Bob. As in \cite{DISTILLATION_LOCCNet}, we specifically focus on the noisy, i.e., mixed, ebit state described by the density matrix
\begin{equation}
\label{eq: S-state}    
    \rho_{A B} = F | \phi^+ \rangle \langle \phi^+ | + (1-F) |00\rangle  \langle 00|,
\end{equation}
where $F \in [0,1]$ represents the \textit{input fidelity} and 
\begin{equation}
\label{eq: phi+ ebit}
    | \phi^+ \rangle = \frac{1}{\sqrt{2}} (|00\rangle + |11\rangle) 
\end{equation}
is a maximally entangled Bell state. The noisy ebit state in (\ref{eq: S-state}) is also known as \textit{S-state} \cite{DISTILLATION_LOCCNet}, and it describes a situation in which the two qubits are in the maximally entangled state, $| \phi^+\rangle$, with probability $F$, and in the separable, i.e., non-entangled, state $|00 \rangle$ with probability $1-F$. This type of noisy state arise in the protocols for entanglement generation that use single-photon detection in the presence of photon loss \cite{Optimizing_practical_entanglement_distillation,Sstate_ref_paper1,Sstate_ref_paper2}. Furthermore, the S state is known to be more challenging to ``denoise" than other mixed states in which the separable state, occurring with probability $1-F$, is orthogonal to $|\phi^+\rangle$ \cite{Optimizing_practical_entanglement_distillation}.

As in \cite{DISTILLATION_LOCCNet}, we focus on the standard case in which $k=2$ identical pairs of S-states $\rho_{A_0 B_0}$ and $\rho_{A_1 B_1}$ are generated. The goal is to \textit{distill} the two noisy ebits pairs to obtain a single pair of less noisy ebits. Following standard terminology \cite{Book_quantum_networking_Meter}, the qubits $A_0$ and $B_0$ are referred to as the \textit{preserved pair}, and the qubits $A_1$ and $B_1$ as the \textit{sacrificial pair}. As shown in Fig. \ref{fig: Alice Bob Charlie}, Alice and Bob process the respective qubits -- $A_1$ and $A_0$ for Alice, and $B_1$ and $B_0$ for Bob -- via \textit{local quantum operations} defined by unitaries $U_A(\theta)$ and $U_B(\theta)$ respectively. As detailed in the next sections, the operation of the unitaries generally depend on a vector $\theta$ of classical parameters. Then, the qubits $A_1$ and $B_1$ are measured in the computational basis at Alice and Bob respectively, and the measurement outcomes ($0$ or $1$) are communicated to Charlie using noisy classical channels. We specifically assume that communication to Charlie occurs over independent \textit{binary symmetric channels} with bit flip probability $p$.

If Charlie receives message $0$ from both Alice and Bob, it declares that the distillation is successful, and Alice and Bob retain the pair of qubits $A_0$ and $B_0$. Instead, Charlie receives the pairs of messages $(0,1),(1,0)$ or $(1,1)$ from Alice and Bob, it declares a failure. In this case, Alice and Bob discard the qubits $A_0$ and $B_0$. 

We remark that most conventional entanglement distillation protocols \cite{DISTILLATION_BBPSW,DISTILLATION_DEJMPS} use decision rules in which either pair of messages $(0,0)$ or $(1,1)$ is considered as success. Here we follow the approach in \cite{DISTILLATION_LOCCNet} of treating $(0,0)$ as the only case in which Charlie declares success. This design choice facilitates the optimization of the unitaries $U_A(\theta)$ and $U_B(\theta)$ through vector $\theta$.

The goal of this work is to design the unitaries $U_A(\theta)$ and $U_B(\theta)$ at Alice and Bob such that the output state of qubits $A_0$ and $B_0$, upon successful distillation, is as close as possible in terms of fidelity to the ideal ebit state $|\phi^+ \rangle$.

\subsection{Performance Metrics}
\label{subsection performance metrics}

The performance of entanglement distillation is measured in this paper, as in \cite{DISTILLATION_LOCCNet,Entanglement_purification_review}, in terms of fidelity and probability of success. The \textit{fidelity} of a state $\rho_{AB}$ with respect to the ebit state $|\phi^+\rangle$ is defined as
\begin{equation}
\label{eq: fidelity definition}
    F(\rho_{AB}) = \langle \phi^+| \rho_{AB} |\phi^+ \rangle ,    
\end{equation}
while \textit{probability of success} is the probability of receiving the pair of messages $(0,0)$ at Charlie.

Let $U(\theta)$ be the $16\times 16$ unitary operation corresponding to the separate application of the $4 \times 4$ local unitaries $U_A(\theta)$ and $U_B(\theta)$ to their respective qubit pairs $(A_0,A_1)$ and $(B_0,B_1)$, respectively. We order the qubits as $(A_0,B_0,A_1,B_1)$ to facilitate the derivations below. The state of the four qubits after the local operations can be expressed as the density matrix
\begin{equation}
\label{eq: rho_out}
    \rho_{out}(\theta) = U(\theta)  (\rho_{A_0 B_0} \otimes \rho_{A_1 B_1})  U(\theta)^{\dagger},
\end{equation}
where we have made explicit dependence on the model parameter vector $\theta$.

The measurement of the sacrificial pair of qubits $(A_1,B_1)$ in the computational basis, $\left\{|00\rangle, |01\rangle, |10\rangle, |11\rangle \right\}$, consists of the projective measurement defined by the four projection matrices
\begin{equation}
\label{eq: POVM}
    \Pi^{xy} =  I_4 \otimes |xy \rangle \langle xy| ,  
\end{equation}
with $(x,y) \in \{0,1\}^2$, where $I_4$ is the $4 \times 4$ density matrix. Accordingly, the measurement returns output $(x,y) \in \{0,1\}^2$ with probability
\begin{equation}
\label{eq: postmeasurement state probability}
    P^{xy} (\theta) = \mathrm{tr}(\Pi^{xy} \rho_{out}(\theta) ),
\end{equation}
and the corresponding post-measurement state
for the qubits $(A_0,B_0)$ is
\begin{equation}
\label{eq: postmeasurement state}
    \rho_{A_0 B_0}^{xy}(\theta) = \frac{(I_4 \otimes \langle xy| ) \rho_{out}(\theta) (I_4 \otimes |xy \rangle) }{P^{xy}(\theta) }.
\end{equation}
Conditioned on the measurement outcome being $(x,y)\in \{0,1\}^2$, the fidelity (\ref{eq: fidelity definition}) of the state $\rho_{A_0 B_0}^{xy}(\theta)$ with respect to the ebit state $|\phi^+\rangle$ is hence
\begin{equation}
\label{eq: fidelity of postmeasurement state}
    F^{xy}(\theta) = \langle \phi^+| \rho_{A_0 B_0}^{xy}(\theta) |\phi^+ \rangle .
\end{equation}

\section{Existing Distillation Protocols}
\label{section literature review}

In this section, we review current state-of-the-art distillation protocols. We focus on the DEJMPS protocol \cite{DISTILLATION_DEJMPS} and on the LOCCNet protocol \cite{DISTILLATION_LOCCNet} as applied to $k=2$ copies of the S-state (\ref{eq: S-state}). We emphasize that all the existing distillation protocols are designed for noiseless classical communication channels to Charlie, i.e., assuming $p=0$.

\subsection{DEJMPS Protocol}
\label{subsection DEJMPS protocols}

In the DEJMPS protocol, the local unitaries $U_A(\theta)$ and $U_B(\theta)$ applied by Alice and Bob do not have free parameters, and are hence denoted as $U_A$ and $U_B$, dropping the dependence on the model parameter vector $\theta$. Specifically, the unitary $U_A$ at Alice is given by Pauli $X$-rotation 
$R_X(\pi/2)$ applied on both qubits, followed by a controlled NOT (CNOT) gate with the qubit $A_0$ as control and the qubit $A_1$ as target. Similarly, the unitary $U_B$ at Bob is defined by the cascade of Pauli $X$-rotations $R_X(-\pi/2)$ on the two qubits and of a CNOT gate with the qubit $B_0$ as control and the qubit $B_1$ as target. If Charlie receives messages $(0,0)$ or $(1,1)$ from Alice and Bob, it declares that distillation is successful, and the qubit pair $(A_0,B_0)$ is retained.

\subsection{LOCCNet}
\label{subsection LOCCNet}

\begin{figure}[htbp]
\centering
\includegraphics[height=2.5in]{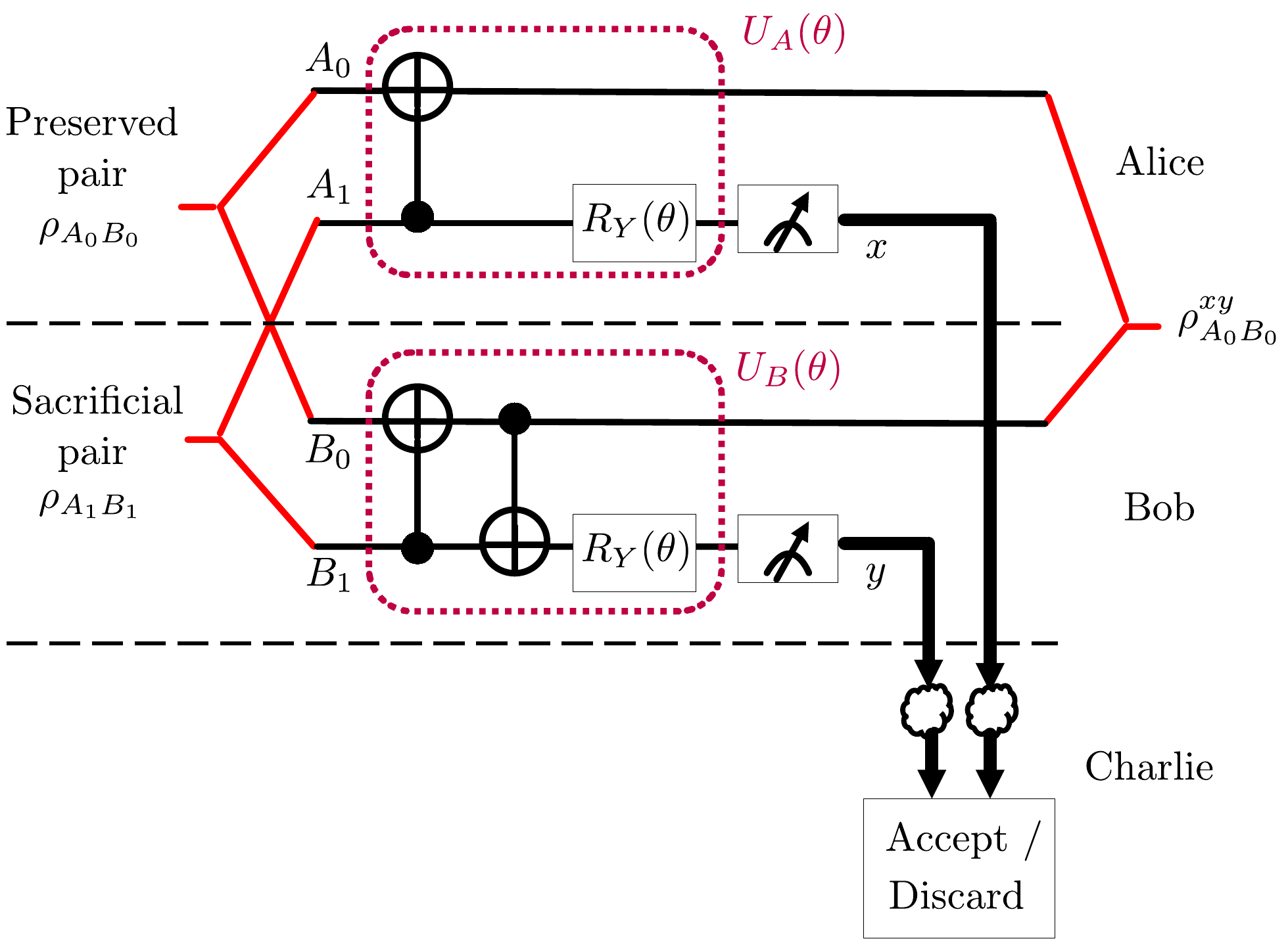}
\caption{LOCCNet circuit for distilling two S states \cite{DISTILLATION_LOCCNet}.}
\label{fig: LOCCNet}
\end{figure}

In \cite{DISTILLATION_LOCCNet}, a quantum machine learning-based entanglement distillation protocol, known as LOCCNet, is introduced that uses parameterized quantum circuits (PQCs) for unitaries $U_A(\theta)$ and $U_B(\theta)$ at Alice and Bob. As illustrated in Fig. \ref{fig: LOCCNet}, the PQC $U_A(\theta)$ consists of a CNOT gate followed by a Pauli $Y$-rotation; while the PQC $U_B(\theta)$ is given by two CNOT gates followed by a Pauli $Y$-rotation. The rotation angle $\theta$ of the Pauli $Y$-rotation is subject to optimization. If Charlie receive messages $(0,0)$ from Alice and Bob through noiseless channels, i.e., $p=0$, a success is declared and the pair $(A_0,B_0)$ of qubits is retained. Model parameter vector $\theta$ is optimized with the goal of maximizing the fidelity $F^{00}(\theta)$ in (\ref{eq: fidelity of postmeasurement state}).

        
\section{Noise Aware-LOCCNet}
\label{section Noisy-LOCCNet}

In this section, we propose \textit{Noise Aware-LOCCNet} (NA-LOCCNet), which distills two qubit pairs, each in the S-state (\ref{eq: S-state}), in the presence of noisy classical channels from Alice and Bob to Charlie as shown in Fig. \ref{fig: Alice Bob Charlie}. The key innovation as compared to LOCCNet is that we explicitly target the performance in terms of average fidelity by accounting for the impact of channel errors. We first describe the design objective, and then introduce the assumed structure for the PQCs $U_A(\theta)$ and $U_B(\theta)$.

\subsection{Design Objective}
\label{subsection design objective}

NA-LOCCNet aims at maximizing the \textit{average conditional fidelity} of a retained pair $(A_0,B_0)$ in case of success. As explained in Section \ref{subsection setting}, Charlie declares a success if it receives the pair of messages $(0,0)$ from Alice and Bob through the respective binary symmetric channels with bit flip probability $p$. LOCCNet assumes a noiseless channel $(p=0)$, and hence it targets the objective $F^{00}(\theta)$, that is, the fidelity conditioned on measurement $(0,0)$ being produced by Alice and Bob. In contrast, NA-LOCCNet accounts for the fact that, where Charlie declares a success as it receives messages $(0,0)$, the actual measurement outcomes may be different due to channel errors.

In fact, messages $(0,0)$ are received at Charlie with probability $(1-p)^2$ if the measurement outcomes are $(x,y) = (0,0)$; with probability $(1-p)p$ if the measurement outcomes are $(x,y) = (0,1)$; with probability $p(1-p)$ if the measurement outcomes are $(x,y) = (1,0)$; and
with probability $p^2$ if the measurement outcomes are $(x,y) = (1,1)$.
Therefore, the average fidelity conditioned on the reception of messages $(0,0)$ is computed as in (\ref{eq: average fidelity for 00 success}), where (\ref{eq: average success probability for 00 case})
\begin{figure*}
\begin{align}
    \label{eq: average fidelity for 00 success}
    \overline{F} (\theta) &=\frac{(1-p)^2 P^{00}(\theta) F^{00}(\theta) + p^2 P^{11}(\theta) F^{11}(\theta) +  p(1-p) P^{10}(\theta) F^{10}(\theta) + (1-p)p P^{01}(\theta) F^{01}(\theta)}{P_{succ}(\theta)} \\ \nonumber \\ 
    \label{eq: average success probability for 00 case}
    P_{succ}(\theta) &= (1-p)^2 P^{00}(\theta) + p^2 P^{11}(\theta) +  p(1-p) P^{10}(\theta) + (1-p)p P^{01}(\theta) 
\end{align}
\end{figure*}
is the probability of success, i.e., of receiving messages $(0,0)$, and we have used definitions (\ref{eq: postmeasurement state probability}) and (\ref{eq: fidelity of postmeasurement state}) (see top of the next page for equations (\ref{eq: average fidelity for 00 success}) and (\ref{eq: average success probability for 00 case})). The proposed protocol NA-LOCCNet addresses the problem
\begin{equation}
\label{eq: optimization expression}
    \underset{\theta}{\max} ~ \overline{F}(\theta).    
\end{equation}

        
\begin{figure}[htbp]
\centering
\includegraphics[height=2.5in]{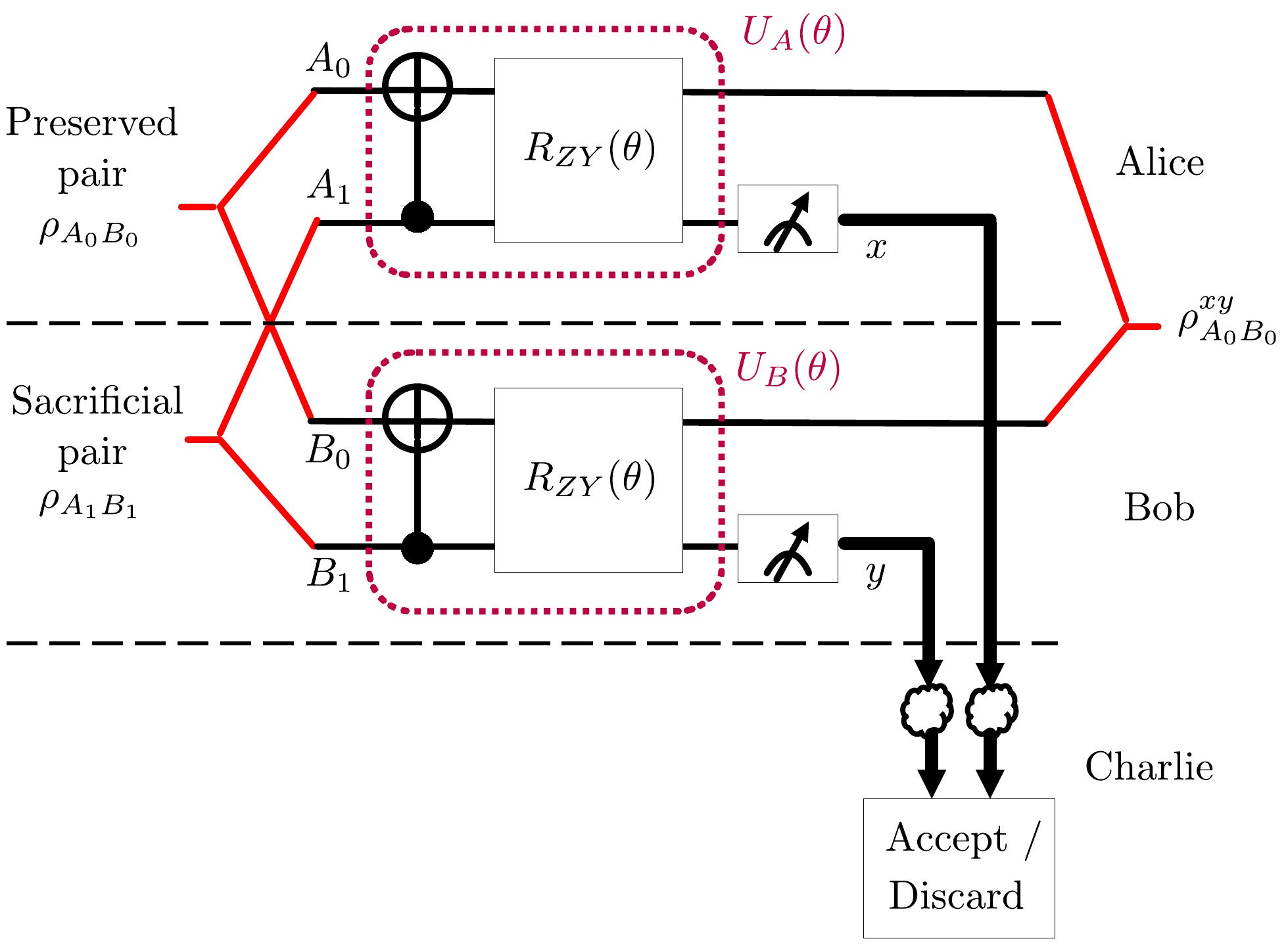}
\caption{Proposed Noise Aware-LOCCNet (NA-LOCCNet) circuit for distilling two S states.}
\label{fig: proposed NA-LOCCNet}
\end{figure}


\subsection{Architecture of the PQCs}
\label{subsectin architecture of the PQCs}

For the PQCs $U_A(\theta)$ and $U_B(\theta)$ at Alice and Bob, respectively, we adopt the architecture shown in Fig. \ref{fig: proposed NA-LOCCNet}. Unlike the LOCCNet architecture in Fig. \ref{fig: LOCCNet}, we introduce a parameterized two-qubit gate, namely the Pauli $ZY$-rotation \cite{Ansatz_TWO_QUBIT_ROTAION_GATES}. This is defined by the unitary
\begin{equation}
\label{eq: two qubit ZY rotation gate}
    R_{ZY}(\theta) = \mathrm{exp}\left(-i \frac{\theta}{2} (Z \otimes Y)\right),
\end{equation}
which is parameterized by angle $\theta$. Recently, two-qubit rotation gates \cite{Ansatz_TWO_QUBIT_ROTAION_GATES} were shown to provide performance advantages as gates in PQCs for various quantum machine learning applications. In our work, the choice of the parameterized two-qubit gate (\ref{eq: two qubit ZY rotation gate}) was dictated by extensive experiments with alternative architectures. As an example, in Section \ref{section experiments}, we will compare the performance obtained by the architecture in Fig. \ref{fig: proposed NA-LOCCNet} with the original LOCCNet system in Fig. \ref{fig: LOCCNet}, when addressing problem (\ref{eq: optimization expression}).

\subsection{Optimization}
\label{subsection optimization}

Addressing problem (\ref{eq: optimization expression}) with PQCs characterized by a single scalar parameter $\theta$, as for the architectures in Fig. \ref{fig: LOCCNet} and Fig. \ref{fig: proposed NA-LOCCNet}, requires a one-dimensional search over the limited domain $[0,2\pi)$. This can be carried out using standard optimization techniques, including grid search or gradient descent.

\section{Experiments}
\label{section experiments}

In this section, we evaluate the performance of the proposed NA-LOCCNet protocol in the presence of noisy communication channels from Alice and Bob to Charlie\footnote{We simulated all the experiments on a laptop with i7 processor and $16$ GB RAM. The PyTorch code for regenerating the results of this paper is available at $<$https://github.com/kclip/Noise-Aware-LOCCNet$>$.}. We consider the benchmark schemes DEJMPS (Section \ref{subsection DEJMPS protocols}) and LOCCNet (Section \ref{subsection LOCCNet}). For the latter, we consider two designs: the original optimization in \cite{DISTILLATION_LOCCNet} of the fidelity $F^{00}(\theta)$ in (\ref{eq: fidelity of postmeasurement state}) and the optimization of the conditional average fidelity $\overline{F}(\theta)$ in (\ref{eq: average fidelity for 00 success}) for the PQC architecture in Fig. \ref{fig: LOCCNet}.


\begin{figure}[htbp]
    \centering
    \includegraphics[height=2.7in]{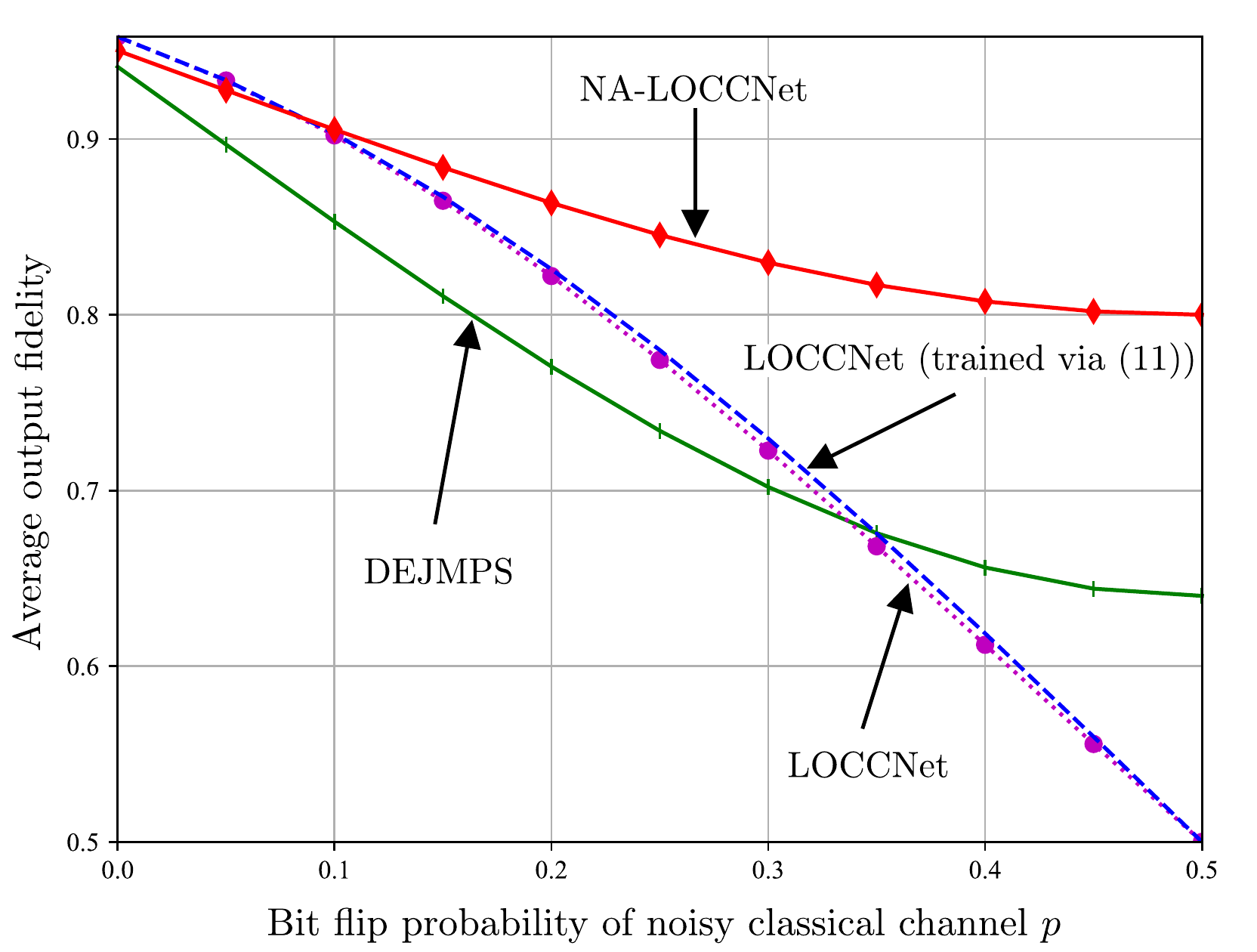}
    \caption{Average output fidelity as a function of the bit flip probability $p$ of the noisy classical channels from Alice and Bob to Charlie for input fidelity $F=0.6$ in (\ref{eq: S-state}).}
    \label{fig: Fig4}
    \vspace{-0.2cm}
\end{figure}



\begin{figure}[htbp]
    \centering
    \includegraphics[height=2.7in]{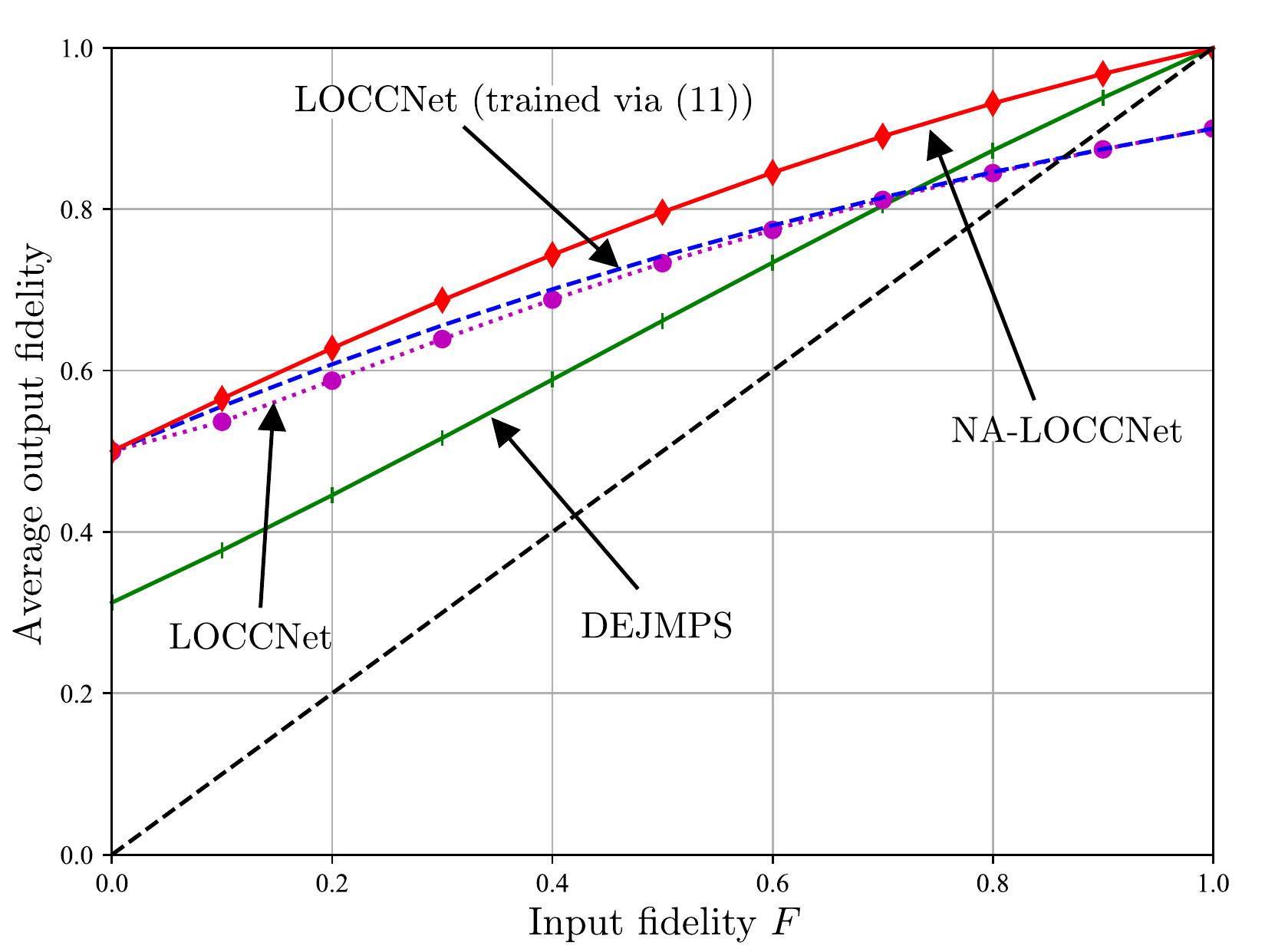}
    \caption{Average output fidelity, conditioned on a successful distillation, as a function of the input fidelity $F$ in (\ref{eq: S-state}) for bit flip probability $p=0.25$ on the noisy classical channels from Alice and Bob to Charlie. The black dashed line corresponds to the reference performance of a scheme that simply outputs the input state.}
    \label{fig: Fig5}
    \vspace{-0.5cm}
\end{figure}



\begin{figure}[htbp]
    \centering
    \includegraphics[height=2.7in]{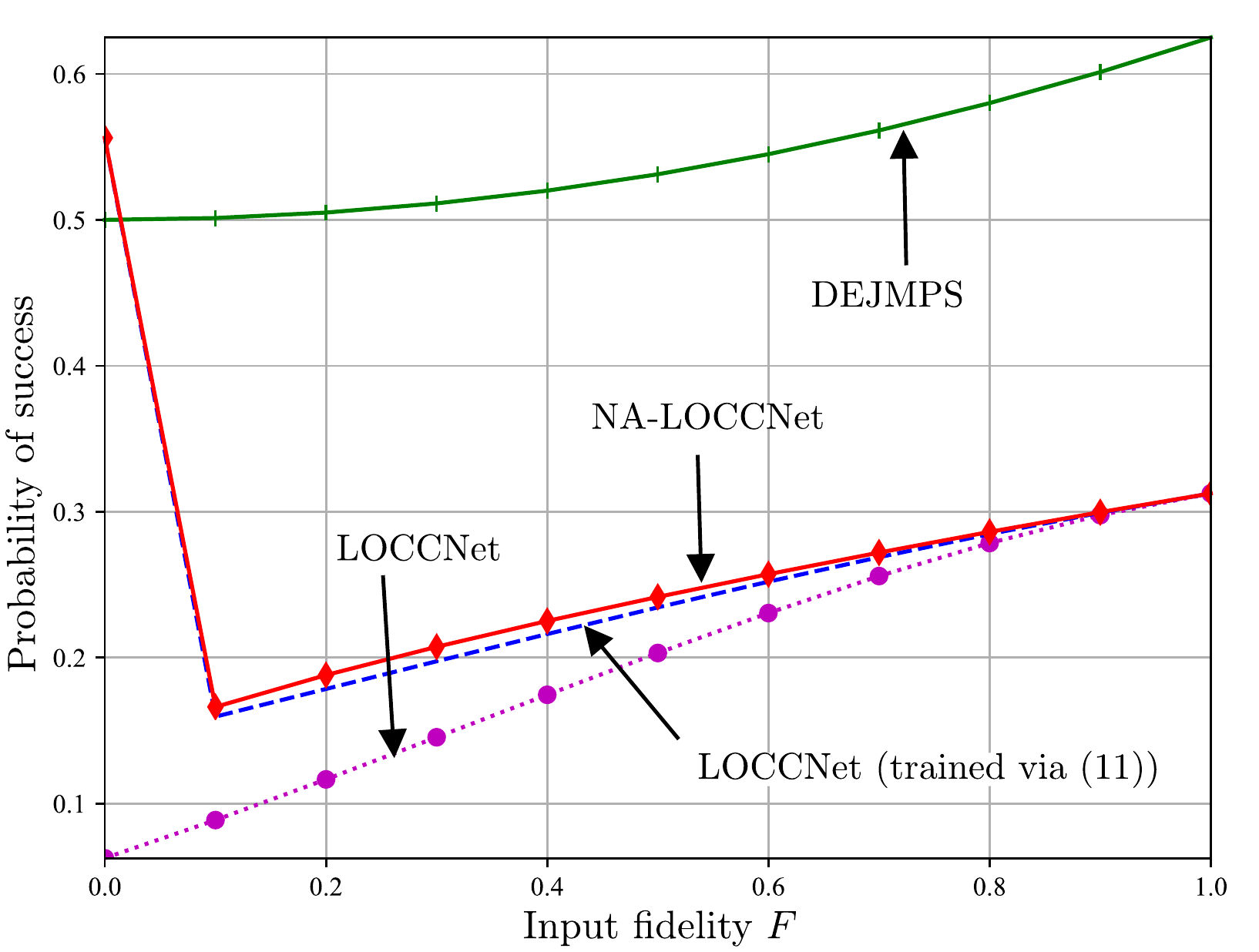}
    \caption{Probability of success as a function of the input fidelity $F$ in (\ref{eq: S-state}) for bit flip probability $p=0.25$ on the noisy classical channels from Alice and Bob to Charlie.}
    \label{fig: Fig6}
    \vspace{-0.5cm}
\end{figure}


Fig. \ref{fig: Fig4} plots the average output fidelity, conditioned on a successful distillation, as a function of the bit flip probability $p$ of the noisy classical channels by fixing the input fidelity of the S-state (\ref{eq: S-state}) to $F=0.6$; while Fig. \ref{fig: Fig5} plots the same quantity as a function of the input fidelity $F$ by fixing the bit flip probability to $p=0.25$. Note that the conditional average fidelity is given by (\ref{eq: average fidelity for 00 success}) for LOCCNet and NA-LOCCNet, while for DEJMPS one needs to consider both received messages $(0,0)$ and $(1,1)$ as indicating successful distillation.

Fig. \ref{fig: Fig4} shows that, as the bit flip probability $p$ increases, the average fidelity of both DEJMPS and LOCCNet decreases significantly, reaching the minimum fidelity of $0.5$ when the channels are maximally noisy, i.e., with $p=0.5$. Note that this fidelity level is smaller than the input fidelity $F=0.6$. Interestingly, the performance of the LOCCNet architecture in Fig. \ref{fig: LOCCNet} does not improve noticeably when optimized via the channel-aware criterion (\ref{eq: optimization expression}), as opposed to the noise-agnostic fidelity criterion considered in \cite{DISTILLATION_LOCCNet}. In contrast, the proposed NA-LOCCNet with PQC architecture in Fig. \ref{fig: proposed NA-LOCCNet} exhibits a significantly milder decrease in fidelity as $p$ grows, yielding the average output fidelity level of $F=0.8$ for $p=0.5$. 

The advantages of NA-LOCCNet are further validated by Fig. \ref{fig: Fig5}, which shows gains at all values of the input fidelity $F$. In particular, unlike the other schemes, NA-LOCCNet never yields an output fidelity lower than the input fidelity $F$.

It is finally noted that the proposed approach, as well as LOCCNet \cite{DISTILLATION_LOCCNet}, target the fidelity performance and not the probability of success. This point is illustrated in Fig. \ref{fig: Fig6}, which shows the probability of success -- given by (\ref{eq: average success probability for 00 case}) for LOCCNet and NA-LOCCNet and by the sum of the probabilities for receiving the messages (0,0) and (1,1) at Charlie for DEJMPS -- as a function of the input fidelity $F$ for $p=0.25$. Overall, NA-LOCCNet is seen to offer a comparable probability of success as compared to LOCCNet, while improving the average fidelity.

\section{Conclusion}
\label{section conclusion}

In this paper, we have studied the problem of entanglement distillation in the presence of noisy classical communications. Specifically, we have proposed to train parameterized quantum circuits (PQCs) at the two parties sharing noisy entangled qubits, so as to maximize the average output fidelity. Unlike existing local operations and classical communication (LOCC) protocols for distillation, which assume ideal noiseless classical communications, the proposed noise aware-LOCCNet (NA-LOCCNet) directly accounts for the presence of noisy channels. Simulation results have confirmed the advantages of the proposed solution.

\bibliographystyle{IEEEtran}
\bibliography{IEEEabrv,cite.bib}

\begin{thebibliography}{10}
\providecommand{\url}[1]{#1}
\csname url@samestyle\endcsname
\providecommand{\newblock}{\relax}
\providecommand{\bibinfo}[2]{#2}
\providecommand{\BIBentrySTDinterwordspacing}{\spaceskip=0pt\relax}
\providecommand{\BIBentryALTinterwordstretchfactor}{4}
\providecommand{\BIBentryALTinterwordspacing}{\spaceskip=\fontdimen2\font plus
\BIBentryALTinterwordstretchfactor\fontdimen3\font minus
  \fontdimen4\font\relax}
\providecommand{\BIBforeignlanguage}[2]{{%
\expandafter\ifx\csname l@#1\endcsname\relax
\typeout{** WARNING: IEEEtran.bst: No hyphenation pattern has been}%
\typeout{** loaded for the language `#1'. Using the pattern for}%
\typeout{** the default language instead.}%
\else
\language=\csname l@#1\endcsname
\fi
#2}}
\providecommand{\BIBdecl}{\relax}
\BIBdecl

\bibitem{Book_Quantum_internet_second_quantum_revolution_2021}
P.~Rohde, \emph{The Quantum Internet: The Second Quantum Revolution}.\hskip 1em
  plus 0.5em minus 0.4em\relax Cambridge University Press, 2021.

\bibitem{Book_quantum_networking_Meter}
R.~Van~Meter, \emph{Quantum Networking}.\hskip 1em plus 0.5em minus 0.4em\relax
  Wiley-IEEE Press, 2014.

\bibitem{Quantum_Internet_IEEE_Network_2020}
A.~S. Cacciapuoti, M.~Caleffi, F.~Tafuri, F.~S. Cataliotti, S.~Gherardini, and
  G.~Bianchi, ``Quantum internet: Networking challenges in distributed quantum
  computing,'' \emph{IEEE Network}, vol.~34, pp. 137--143, 2020.

\bibitem{Book_quantum_computing_quantum_information_Nielsen_chuang_2010}
M.~A. Nielsen and I.~L. Chuang, \emph{Quantum Computation and Quantum
  Information}.\hskip 1em plus 0.5em minus 0.4em\relax Cambridge University
  Press, 2010.

\bibitem{Book_Quantum_Information_Theory_wilde_2013}
M.~M. Wilde, \emph{Quantum Information Theory}.\hskip 1em plus 0.5em minus
  0.4em\relax Cambridge University Press, 2013.

\bibitem{single_photon_sources_and_detectors}
M.~D. Eisaman, J.~Fan, A.~Migdall, and S.~V. Polyakov, ``Invited review
  article: Single-photon sources and detectors,'' \emph{Review of scientific
  instruments}, vol.~82, no.~7, p. 071101, 2011.

\bibitem{Sstate_ref_paper2}
E.~T. Campbell and S.~C. Benjamin, ``Measurement-based entanglement under
  conditions of extreme photon loss,'' \emph{Phys. Rev. Lett.}, vol. 101, p.
  130502, Sep 2008.

\bibitem{book_machine_learning_with_quantum_computers_Maria_Francesco}
M.~Schuld and F.~Petruccione, \emph{Machine Learning with Quantum
  Computers}.\hskip 1em plus 0.5em minus 0.4em\relax Springer Cham, 2021.

\bibitem{Book_Osvaldo_quantum_machine_learning_for_engineers}
O.~Simeone, \emph{An Introduction to Quantum Machine Learning for
  Engineers}.\hskip 1em plus 0.5em minus 0.4em\relax
  https://osimeone.wordpress.com/, 2022.

\bibitem{DISTILLATION_BBPSW}
C.~H. Bennett, G.~Brassard, S.~Popescu, B.~Schumacher, J.~A. Smolin, and W.~K.
  Wootters, ``Purification of noisy entanglement and faithful teleportation via
  noisy channels,'' \emph{Phys. Rev. Lett.}, vol.~76, pp. 722--725, Jan 1996.

\bibitem{DISTILLATION_DEJMPS}
D.~Deutsch, A.~Ekert, R.~Jozsa, C.~Macchiavello, S.~Popescu, and A.~Sanpera,
  ``Quantum privacy amplification and the security of quantum cryptography over
  noisy channels,'' \emph{Phys. Rev. Lett.}, vol.~77, pp. 2818--2821, Sep 1996.

\bibitem{DISTILLATION_LOCCNet}
X.~Zhao, B.~Zhao, Z.~Wang, Z.~Song, and X.~Wang, ``Practical distributed
  quantum information processing with loccnet,'' \emph{Quantum Information},
  vol.~7, no.~1, pp. 1--7, 2021.

\bibitem{Optimizing_practical_entanglement_distillation}
F.~Rozpedek, T.~Schiet, L.~P. Thinh, D.~Elkouss, A.~C. Doherty, and S.~Wehner,
  ``Optimizing practical entanglement distillation,'' \emph{Phys. Rev. A},
  vol.~97, p. 062333, Jun 2018.

\bibitem{Sstate_ref_paper1}
N.~H. Nickerson, J.~F. Fitzsimons, and S.~C. Benjamin, ``Freely scalable
  quantum technologies using cells of 5-to-50 qubits with very lossy and noisy
  photonic links,'' \emph{Phys. Rev. X}, vol.~4, p. 041041, Dec 2014.

\bibitem{Entanglement_purification_review}
W.~Dür and H.~J. Briegel, ``Entanglement purification and quantum error
  correction,'' \emph{Reports on Progress in Physics}, vol.~70, no.~8, pp.
  1381--1424, jul 2007.

\bibitem{Ansatz_TWO_QUBIT_ROTAION_GATES}
J.-B. You, D.~E. Koh, J.~F. Kong, W.-J. Ding, C.~E. Png, and L.~Wu, ``Exploring
  variational quantum eigensolver ansatzes for the long-range xy model,''
  \emph{arXiv preprint arXiv:2109.00288}, 2021.

\end{thebibliography}

\end{document}